\documentclass[twocolumn]{aastex62}

\usepackage{amsmath}
\usepackage{amssymb}
\usepackage{CJK}
\usepackage{multirow} 
\usepackage{ulem}
\usepackage{color}

\begin{document}
\begin{CJK*}{UTF8}{gbsn}
\title{M17~MIR: A Massive Star Is Forming via Episodic Mass Accretion}
\correspondingauthor{Zhiwei Chen}
   \email{zwchen@pmo.ac.cn}

\author[0009-0001-5449-1939]{Wei Zhou (周威)}
\affiliation{Purple Mountain Observatory, Chinese Academy of Sciences, Nanjing 210023, People's Republic of China}
\affiliation{University of Science and Technology of China, Chinese Academy of Sciences, Hefei 230026, People's Republic of China}

\author[0000-0003-0849-0692]{Zhiwei Chen (陈志维)}
\affiliation{Purple Mountain Observatory, Chinese Academy of Sciences, Nanjing 210023, People's Republic of China}

\author[0000-0002-5920-031X]{Zhibo Jiang (江治波)}
\affiliation{Purple Mountain Observatory, Chinese Academy of Sciences, Nanjing 210023, People's Republic of China}

\author[0000-0003-1714-0600]{Haoran Feng (冯浩然)}
\affiliation{Purple Mountain Observatory, Chinese Academy of Sciences, Nanjing 210023, People's Republic of China}
\affiliation{University of Science and Technology of China, Chinese Academy of Sciences, Hefei 230026, People's Republic of China}

\author[0000-0002-3549-5029]{Yu Jiang (蒋禹)}
\affiliation{Purple Mountain Observatory, Chinese Academy of Sciences, Nanjing 210023, People's Republic of China}
\affiliation{Center for Astronomy and Space Sciences, China Three Gorges University,  Yichang 443002, People's Republic of China}

\begin{abstract}
We analyzed the Atacama Large Millimeter/submillimeter Array (ALMA) band 6 data for the outbursting massive protostar M17~MIR. The ALMA CO $J=2-1$ data reveal a collimated and bipolar north–south outflow from M17~MIR. The blueshifted outflow exhibits four CO knots (N1 to N4) along the outflow axis, while the redshifted outflow appears as a single knot (S1). The extremely high velocity (EHV) emissions of N1 and S1 are jetlike and contain subknots along the outflow axis. Assuming the nearest EHV subknots trace the ejecta from the accretion outbursts in the past decades, a tangential ejection velocity of $\sim421\,\mathrm{km\,s^{-1}}$ is derived for M17~MIR.  Assuming the same velocity, the dynamical times of the multiple ejecta, traced by the four blueshifted CO knots, range from 20 to 364 yr. The four blueshifted CO knots imply four clustered accretion outbursts with a duration of tens of years in the past few hundred years. The intervals between the four clustered accretion outbursts are also about tens of years. These properties of the four clustered accretion outbursts are in line with the disk gravitational instability and fragmentation model. The episodic accretion history of M17~MIR traced by episodic outflow suggests that a massive star can form from a lower-mass protostar via frequent episodic accretion events triggered by disk gravitational instability and fragmentation. The first detection of the knotty outflow from an outbursting massive protostar suggests that mass ejections accompanied with accretion events could serve as an effective diagnostic tool for the episodic accretion histories of massive protostars.
\end{abstract}

\keywords{Star formation (1569); Jet outflows (1607); Stellar accretion (1578)}

\section{Introduction} \label{sec:intro}
The formation mechanisms of massive young stellar objects (MYSOs) are still far from being fully understood. Disk accretion models and observational evidence suggest that the disk accretion at a high rate, $\gtrsim 10^{-4}\,M_\sun\,\mathrm{yr}^{-1}$, is vital for the formation of MYSOs  \citep{2016A&ARv..24....6B}. The newly discovered accretion-bursting MYSOs suggest that episodic accretion events might be important phases of the formation process of massive stars \citep{2017NatPh..13..276C,2017ApJ...837L..29H,2021A&A...646A.161S,2021ApJ...922...90C}. 

Besides the infrared and maser variability accompanying accretion outbursts, molecular outflows launched during accretion outbursts could be recorded in the form of separated outflow knots or bullets along the outflow axis \citep{2009ApJ...702L..66Q,2016ApJ...824...72C,2024ApJ...961..108K}. 
The separation between various outflow knots could be used to constrain the time interval between each individual accretion outburst \citep{2018A&A...613A..18V,2022MNRAS.510.2552R}. The time interval between accretion outbursts, depending on the triggering mechanism, varies from tens to thousands of years \citep{2019ApJ...872..183F,2019MNRAS.482.5459M,2021A&A...651L...3E}. Currently, infrared and maser light curves only cover a minor fraction of the main accretion phase of MYSOs, preventing a detailed understanding of the episodic accretions of MYSOs over a long period. The outflow dynamical time of low-mass young stars can be up to thousands of years \citep{2015Natur.527...70P, 2022ApJ...924...50B}.
 To date, little is known about the features of molecular outflows powered by outbursting MYSOs and their relation with the accretion processes. 

To date, only few MYSOs, S255IR~NIRS~3 \citep{2017NatPh..13..276C}, NGC~6334I MM1 \citep{2017ApJ...837L..29H}, G358.93--0.03 MM1 \citep{2021A&A...646A.161S}, and M17~MIR \citep{2021ApJ...922...90C}, have been found to exhibit accretion outbursts.  M17~MIR, located at $\alpha_\mathrm{J2000}=18^\mathrm{h}20^\mathrm{m}23\fs017$, $\delta_\mathrm{J2000}=-16\degr 11\arcmin 47\farcs98$, is an extremely red protostar embedded within the massive star-forming cloud M17~SW at a distance $\approx 2.0\,\mathrm{kpc}$ \citep{2011ApJ...733...25X,2016MNRAS.460.1839C}. The infrared light curve of M17~MIR suggests two infrared luminosity bursts and a quiescent phase in between. The major one occurred in the 1990s, while the ongoing one started in mid-2010. The H$_2$O maser variability of M17~MIR is roughly contemporaneous with the major luminosity burst in the 1990s
 \citep[\citealt{1989A&A...213..339F,1998ApJ...500..302J,2010MNRAS.406.1487B}; summarized in Table 6 of][]{2021ApJ...922...90C}. The 3D motions of the H$_2$O maser spots, measured during the transient from the quiescent phase to the ongoing luminosity burst, show an expanding bubble structure originating from M17~MIR \citep{2016MNRAS.460.1839C}. The infrared luminosity bursts and H$_2$O maser variability suggest that M17~MIR is a unique MYSO with recurrent accretion outbursts, one in the 1990s with $\dot M_\mathrm{acc}\sim5\times10^{-3}\,M_\sun\,\mathrm{yr}^{-1}$ and an ongoing one with $\dot M_\mathrm{acc}\sim1.7\times10^{-3}\,M_\sun\,\mathrm{yr}^{-1}$. The ongoing accretion outburst leads to a luminosity burst of $\sim 7600\,L_\sun$. Although the current mass of M17~MIR is estimated to be $\sim5.4\,M_\sun$, the final mass of M17~MIR could potentially reach $\gtrsim20\,M_\sun$, equivalent to an O9 type star  \citep{2021ApJ...922...90C}. In this work, we present the first detection of episodic mass ejection from M17~MIR, based on the millimeter interferometric observations taken with the Atacama Large Millimeter/submillimeter Array (ALMA) in band 6.

\section{Archival ALMA band 6 data} \label{sec:data}
The ALMA band 6 observations (project ID: 2019.1.00 \\ 994.S) of M17~MIR were conducted with the 12\,m array on 2019 October 20 during ALMA Cycle 6. The ALMA band 6 observations covered four spectral windows (SPWs; SPW 1: $219.428-219.662$ GHz, SPW 2: $220.266-220.5$ GHz, SPW 3: $230.405-230.639$ GHz, and SPW 4: $232.041-233.915$ GHz).  The data reduction was done by the Science Ready Data Products (SRDP) Initiative \footnote{https://science.nrao.edu/srdp} \citep{2020ASPC..527..519L}, developed by the National Radio Astronomy Observatory. The $^{12}$CO $J=2-1$ transition at 230.538 GHz is covered by the SPW 3 data with a velocity width of $0.32\,\mathrm{km\,s^{-1}}$ per channel and an rms noise of $\sim4$ mJy beam$^{-1}$ per channel. The synthesized beam of the SPW 3 data is $0\farcs705\times0\farcs452$. SRDP also returned the 1.3 mm continuum map, which was produced by averaging all the line-free channels of the data in SPWs $1-4$. The synthesized beam size of the 1.3 mm continuum is $0\farcs725\times0\farcs441$, and the rms noise is $\sim0.7$ mJy beam$^{-1}$.

\section{Results} \label{sec:result}

\begin{figure*}[ht!]
\centering
\includegraphics[width=\textwidth]{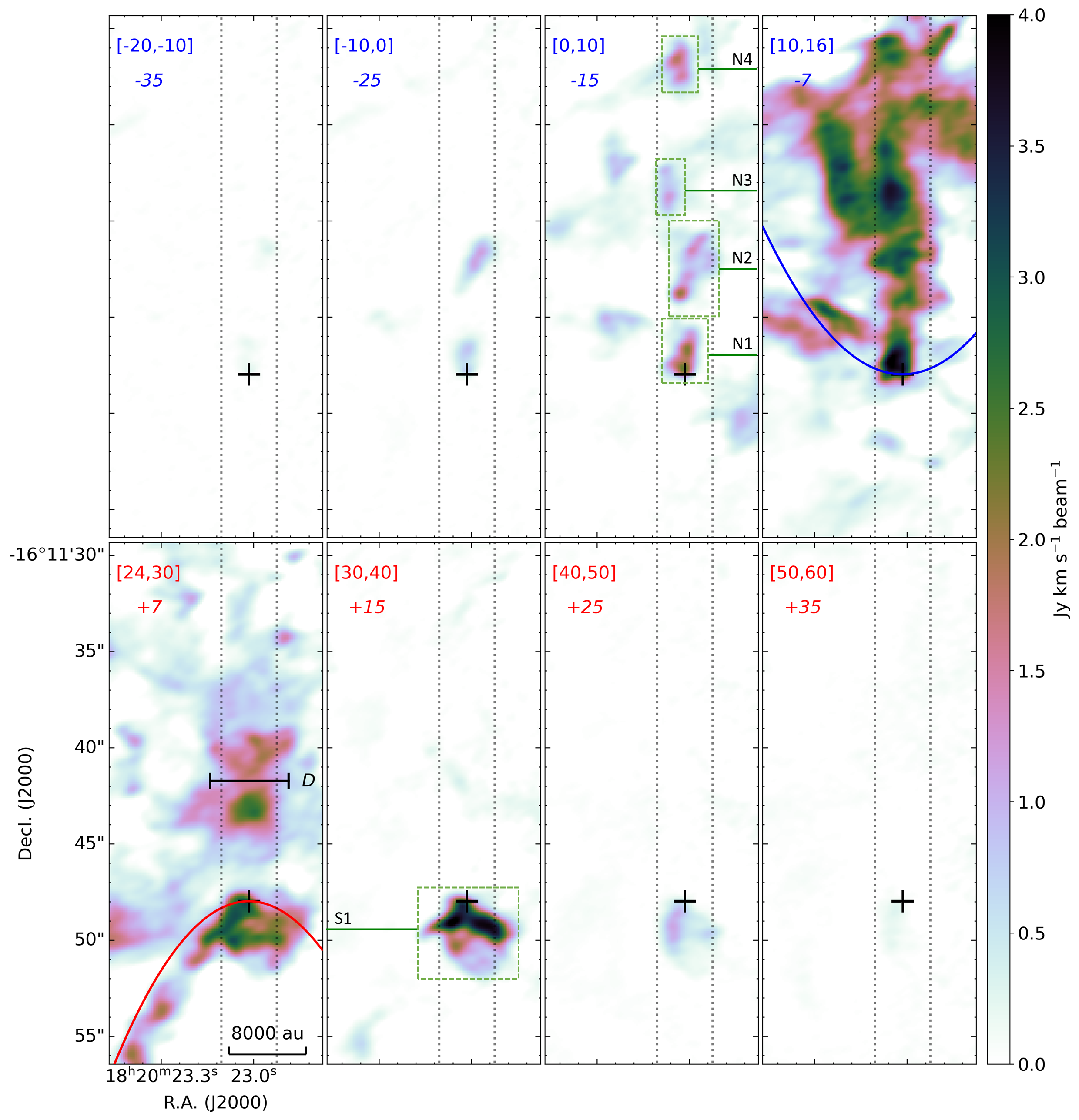}
\caption{ Channel maps of $^{12}$CO $J=2-1$  for the eight velocity ranges from $-20$ to $60\,\mathrm{km\,s^{-1}}$. The velocity range and the relative velocity to $V_\mathrm{sys}$ are shown in the upper left corner of each map. The red and blue solid lines illustrate wide parabolic cavity walls carved by the red- and blueshifted outflow, respectively. The black horizontal line in the $24-30\,\mathrm{km\,s^{-1}}$ channel represents the extension of N2 perpendicular to the outflow axis. The two vertical dotted lines $3\arcsec$ apart outline the areas for generating the PV diagram in Figure~\ref{fig:pv_diagram}(a). The outflow knots are marked within the dashed rectangle. The black cross indicates the position of M17~MIR.}
\label{fig:ch_map}
\end{figure*}

\begin{figure*}[ht!]
\centering
\includegraphics[width=\textwidth]{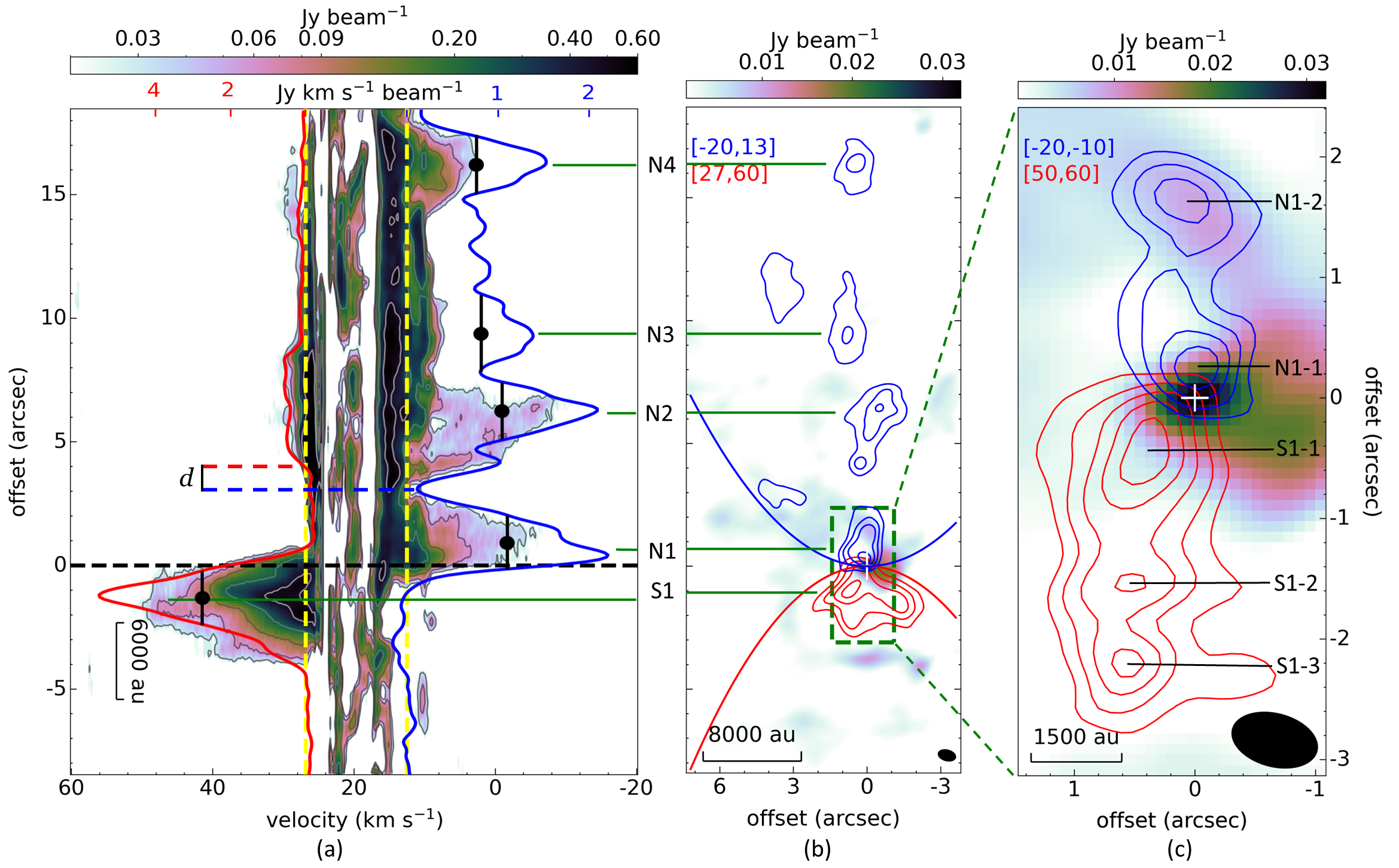}
\caption{(a) PV diagram of the CO $J=2-1$ line emission along the outflow axis. The yellow vertical dashed lines indicate the velocity range $13-27\,\mathrm{km\,s^{-1}}$. The black horizontal dashed line denotes the position of M17~MIR on the slice. The red and blue solid curves represent the integrated intensity in the slice with $3\arcsec$ width varying with the slice offset. The \textit{d} represents the axial offset between the blue- and redshifted components of the N2 knot. The positions of the CO knots, from which the distances of the CO knots to the central source are measured, are marked by the black dots overlaid on the vertical black segments (the lengths of the CO knots). (b) Integrated intensity of the red- and blueshifted CO $J=2-1$ emission with $|V-V_{\mathrm{sys}}|>7\,\mathrm{km\,s^{-1}}$ are displayed as the red and blue contours, respectively, overlaid on the ALMA 1.3 mm continuum map. The synthesized beam of the 1.3 mm continuum is drawn in the lower right corner. The red/blue contours start from 30\% of the red-/blueshifted integrated emission peak ($\sim\mathrm{1.6\,/\,2.2\,  Jy\,km\,\,s^{-1}\,beam^{-1}}$), with steps of 20\% of the red/blue integrated emission peak. The white cross indicates the position of M17~MIR. (c) Zoomed-in map of Figure~\ref{fig:pv_diagram}(b), depicting the EHV components of the red- and blueshifted outflows ($|V-V_{\mathrm{sys}}|>30\,\mathrm{km\,s^{-1}}$). The blue and red contours both start from $5\sigma$ level with a step of $3\sigma$ level, where $\sigma\approx15\,\mathrm{mJy\,km\,s^{-1}\,beam^{-1}}$.}
\label{fig:pv_diagram}
\end{figure*}

M17~MIR is located at the M17~SW cloud which is just adjacent to the \ion{H}{2} region M17. The molecular gas of M17~SW is highly disturbed and strongly heated by the \ion{H}{2} region \citep{2010A&A...510A..87P,2012A&A...542L..13P,2015A&A...583A.107P}. The CO $J=2-1$ line emission averaged over the inner $200\arcsec$ in diameter of the M17~SW cloud peaked at $19.88\,\mathrm{km\,s^{-1}}$ with an FWHM of $8.32\,\mathrm{km\,s^{-1}}$ \citep{2015A&A...583A.107P}. 
 
The class I methanol maser and H$_2$O masers associated with M17~MIR indicate outflow activities \citep{2010ApJS..191..207G,2021ApJ...922...90C}.
To reveal the putative molecular outflow from M17~MIR, we explored the ALMA band 6 data for M17~SW to search for the high-velocity CO gas around M17~MIR. Figure~\ref{fig:ch_map} displays the CO $J=2-1$ channel maps from $-20$ to $60\,\mathrm{km\,s^{-1}}$ around M17~MIR. Assuming a radial velocity of $20\,\mathrm{km\,s^{-1}}$ for M17~MIR, the ALMA data reveal high-velocity CO gas reaching relative velocity $\gtrsim4\,\mathrm{km\,s^{-1}}$ on the northern and southern sides of M17~MIR. The blue- and redshifted CO gas intersects at the position of M17~MIR. The redshifted gas of low velocity ($4<V-V_\mathrm{sys} < 20\,\mathrm{km\,s^{-1}}$) appears to be parabolic south of M17~MIR, reminiscent of the wide cavity wall carved by outflow \citep{2000ApJ...542..925L,2016ARA&A..54..491B}. At higher velocity ($V-V_\mathrm{sys}>20\,\mathrm{km\,s^{-1}}$), collimated jetlike outflow starts to appear. The wide cavity wall carved by the redshifted outflow is illustrated as the red parabolic curve in Figure~\ref{fig:ch_map}. On the opposite side, where the blueshifted CO emission is observed, a similar cavity wall is prominent at low velocity, illustrated by the blue parabolic curve. The blueshifted outflow extends farther to the north and displays distinct knotty structures along the outflow axis.

\begin{table*}[ht!]\footnotesize
\centering
\caption{Observed Properties of the Knotty Structures}
\label{tbl:1}
\begin{tabular}{ccccccc}
\hline\hline
Knot & $d$    & $V_\mathrm{max}$  & $V_\mathrm{avg}$  & $L$   & Subknot  & $d$\\
 & (arcsec) & (km s$^{-1}$) & (km s$^{-1}$) & (arcsec) &  & (arcsec)\\
\hline
\multirow{2}{*}{N1}   &  \multirow{2}{*}{0.9} & \multirow{2}{*}{33 (115)} & \multirow{2}{*}{13 (45)}  & \multirow{2}{*}{2.3} & 1 &  0.4\\
& & & & & 2 & 1.5 \\
 \hline
N2   &  6.2 & 32 (112)  & 13 (45) & 2.4 & &  \\
N3   &  9.4  & 13 (45) & 8  (28) & 3.0 & &   \\
N4   &  16.2 & 21 (73)& 10 (35) & 2.5 & & \\
\hline
\multirow{3}{*}{S1}  &  \multirow{3}{*}{1.3} & \multirow{3}{*}{37 (129)} & \multirow{3}{*}{13 (45)} & \multirow{3}{*}{2.3} & 1 & 0.5\\
& &&&& 2& 1.6\\
& &&&& 3& 2.2\\
\hline 
\end{tabular}
\end{table*}

Chains of elongated knots are prominently observed to the north of M17~MIR. These knots are also distinctly visible in the position–velocity (PV) diagram along the outflow axis (Figure~\ref{fig:pv_diagram}(a)). Four major outflow knots of the blueshifted outflow are separated along the outflow axis and are referred to as N1, N2, N3, and N4. The redshifted outflow is referred to as S1 since only one major peak is seen in the PV diagram. For $|V-V_{\mathrm{sys}}|\geqslant7\,\mathrm{km\,s^{-1}}$, the low-velocity emission from the cloud is mostly excluded, whereas emission from knotty outflow structure becomes prominent. The integrated CO emission with velocities $|V-V_{\mathrm{sys}}|\geqslant7\,\mathrm{km\,s^{-1}}$ is shown as the red and blue contours in Figure~\ref{fig:pv_diagram}(b). Similar to the channel maps, the blueshifted knots N1 to N4 are distributed along the outflow axis, and the redshifted knot S1 shows a wide-angle cavity morphology.

The nearest knot N1 is elongated along the outflow axis. The close-up view of N1 at extremely high velocity (EHV; $|V-V_{\mathrm{sys}}|>30\,\mathrm{km\,s^{-1}}$) is displayed in Figure~\ref{fig:pv_diagram}(c). These EHV components of N1 present two distinguished subknots, N1-1 and N1-2. The EHV components of S1 also include subknots along the outflow axis, contributing to the appearance of a symmetric jetlike structure when combined with the EHV components of N1.
N2 displays faint redshifted emission, suggesting the presence of vertical-splash gas \citep{2022ApJ...931L...5J}. 
N3 and N4 show reduced high-velocity components, possibly due to the prolonged interaction with the surrounding gas. N4 also exhibits faint redshifted emission similar to N2.

For the CO knots revealed by the ALMA band 6 data at 1000 au resolution, we calculated their projected distances ($d$) from M17~MIR, maximum outflow velocities ($V_\mathrm{max}$), mean outflow velocity ($V_\mathrm{avg}$), and projected lengths ($L$) along the outflow axis. The estimation of $V_\mathrm{avg}$ only considers the high-velocity emission ($|V-V_{\mathrm{sys}}|>7\,\mathrm{km\,s^{-1}}$) of each knot to exclude the influence of the natal cloud. The inclination angle of the CO outflow is $\sim16\degr$ (see Appendix~\ref{sec:appen_A} for more details). The tangential velocity of each CO knot is then derived and listed in the brackets of the third and fourth columns of Table~\ref{tbl:1}. In Figure~\ref{fig:pv_diagram}(a), $L$ is denoted by the black segments with the solid dot in the middle, which is defined as the full width at half maximum along the outflow axis with an exception for N3. The $L$ of N3 is the length between the two nearest local minima along the outflow axis. The solid dots represent the positions of the major knots along the outflow axis, from which the projected distances ($d$) are calculated. These observed properties of the major knots are listed in Table~\ref{tbl:1}. The $d$ values for the subknots of N1 and S1 are provided in the rightmost column of Table~\ref{tbl:1}, calculated using the same method as for the major CO knots but only for the EHV CO emission. 

\begin{table}[ht!]
\footnotesize
\centering
\caption{Physical Parameters of the Knotty Structures}
\label{tbl:2}
\begin{tabular}{ccccccc}
\hline\hline
Knot  & $t_\mathrm{dyn}$ & $M/10^{-2}$  & $\Delta t$   & $\dot M_\mathrm{out}/10^{-4}$  & Sub-  & $t_\mathrm{dyn}$\\ 
       & (yr) & ($M_\sun$) & (yr) & ($M_\sun\,\mathrm{yr}^{-1}$)  & knot   & (yr)\\
\hline
\multirow{2}{*}{N1} & \multirow{2}{*}{20} & \multirow{2}{*}{1.6} & \multirow{2}{*}{51}  &\multirow{2}{*}{3.1}   & 1 & 9 \\
& &&& & 2 & 34 \\
\hline
N2 & 140 &2.0 & 53  & 3.7  & & \\
N3 & 213  &1.5 & 68  & 2.2  & &\\
N4 & 364  &1.0 & 55  & 1.8  & &\\
\hline
\multirow{3}{*}{S1} & \multirow{3}{*}{29}  &\multirow{3}{*}{5.1} & \multirow{3}{*}{51}  & \multirow{3}{*}{10.0}   & 1  & 11  \\
& & &&& 2 & 36 \\
& & &&& 3 & 50 \\
\hline 
\end{tabular}
\end{table}

\section{Discussion} \label{sec:discuss}
\subsection{Episodic mass ejection as a consequence of accretion outburst}
Episodic ejecta serves as an indicator of variable or discrete ejection events, presumably linked to unsteady and episodic mass accretion processes \citep{2014A&A...563A..87E,2015ApJ...805..115V,2018A&A...613A..18V}. 
The radio jet, launched by the recent accretion event of the outbursting MYSO S255IR~NIRS3, was observed to expand at a speed of $\sim285\,\mathrm{km\,s^{-1}}$ \citep{2024A&A...683L..L15C}. In comparison, jet velocities of other infrared bright MYSOs, measured from shock gas tracers in the near-infrared, range from $\sim80$ to $450\,\mathrm{km\,s^{-1}}$ \citep{2018A&A...616A.126F,2023A&A...676A.107F,2023A&A...672A.113M}.
Currently, neither radio continuum nor near-infrared high-resolution data from multiple epochs are available for M17~MIR. Therefore, the knotty structures of the north–south CO outflow, driven by the outbursting MYSO M17~MIR, are crucial for establishing the connection between episodic ejecta and accretion events in the early stages of MYSOs.

M17~MIR underwent a major accretion outburst around the 1990s and is currently undergoing its second moderate accretion outburst started in mid-2010 \citep{2021ApJ...922...90C}. 
Interestingly, we observe symmetrical EHV subknots very close to the central sources on both red- and blueshifted sides, suggesting symmetrical mass ejections on both sides likely triggered by the accretion outbursts reported in \citet{2021ApJ...922...90C}.  
Similar evidence of a connection between mass ejection and accretion outburst has also been observed for the low-mass protostar B335 \citep{2024ApJ...961..108K} and MYSO S255IR~NIRS~3 \citep{2023A&A...676A.107F,2023A&A...680A.110C}. It is reasonable to assume that the closest EHV subknots N1-1 and S1-1 represent the mass ejection caused by the ongoing accretion outburst, which started 9 yr before the ALMA observations in 2019 October. Therefore, the dynamical time of the EHV subknots N1-1 and S1-1 is restricted to $\sim 9$ yr. The tangential velocity of the ejecta triggered by the ongoing accretion outburst is then derived to be $\sim421\,\mathrm{km\,s^{-1}}$. 

The second closest EHV subknots N1-2 and S1-2 are assumed to be related to the major accretion outburst around the 1990s although the beginning phase of this earlier accretion outburst is poorly constrained due to the lack of mid-IR light curve data before the 1990s. We assume that the tangential ejection velocity triggered by this earlier accretion outburst around the 1990s is $\sim421\,\mathrm{km\,s^{-1}}$, the same as for the ongoing accretion outburst. The dynamic time of the ejecta traced by N1-2 and S1-2 is constrained to $d/(421\,\mathrm{km\,s^{-1}})\approx35\,\mathrm{yr}$, where $d$ is the mean distance of the EHV subknots N1-2 and S1-2 to the central source. A dynamical time of 35 years agrees fairly well with that of the major accretion outburst that likely began between 1984 and 1993 \citep{2021ApJ...922...90C}, an implication that the tangential ejection velocity assumed for the EHV subknots N1-2 and S1-2 is reasonable. Furthermore, the assumed tangential ejection velocity is much higher than the tangential maximum and mean velocities of the CO knots (see Table~\ref{tbl:1}), in agreement with the expectation that the molecular outflow is the entrained secondary envelope gas \citep{2017A&A...607L...6T,2019ApJ...883....1Z}.

The farthest EHV subknot S1-3 might imply an accretion outburst occurred even earlier than the recent two accretion outbursts. With the assumption of the same ejection velocity as other subknots, the dynamical time of this potential accretion outburst is $\sim 50\,\mathrm{yr}$, longer than the time span of mid-IR observations available for M17~MIR. The more distant CO knots, N2 to N4, might trace mass ejections with much longer dynamical times. Assuming the mass ejections traced by these distant CO knots move at a constant tangential velocity of $\sim421\,\mathrm{km\,s^{-1}}$, we derive their dynamical times, which are presented in the second column of Table~\ref{tbl:2}. The dynamical time increases from $140$ to $364$ yr for the CO knots N2 to N4 from near to far. Because the velocity of ejecta might be slowed down due to the interaction with the ambient gas \citep[e.g.][]{2024A&A...683L..L15C}, the dynamical times of the CO knots N2 to N4 provide us the lower limits on the timescales of the ejecta traced by them. The farthest CO knot N4 likely represents the oldest ejecta with a timescale longer than $364$ yr.

The disk gravitational instability and fragmentation model predicts two types of accretion outbursts: the isolated ones with $10^4$ yr periods and clustered ones containing several bursting events one after another during just tens of years \citep{2015ApJ...805..115V}. The EHV subknots of the CO knots N1 and S1 suggest three accretion outbursts of M17~MIR in the past 50 yr, the most recent two accretion outbursts were confirmed by the bursts of both mid-IR luminosity and H$_2$O maser emission in the past 40 yr \citep{2021ApJ...922...90C} and a presumably third accretion outburst with a timescale of 50 yr (relative to the ALMA observations in 2019), which are in line with the clustered type of accretion outbursts predicted by \citet{2015ApJ...805..115V}. The four blueshifted CO knots N1 to N4 distributed along the outflow axis imply the four clustered accretion outbursts in the formation history of M17~MIR \citep[e.g.][]{2018A&A...613A..18V}. The $t_\mathrm{dyn}$ of the CO knots are treated as the coarse approximation to the middle ages of the ejecta triggered by the clustered accretion outbursts. In contrast to the infrared observations conducted only in recent decades, high-resolution ALMA observations are capable of revealing mass ejection occurred up to hundreds of years ago. The earliest accretion outburst cluster, traced by knot N4, might have occurred several hundred years ago.

We also estimate the duration ($\Delta t$) of each clustered accretion outburst by dividing the length ($L$) of each CO knot by the previously assumed tangential ejection velocity of $\sim 421\,\mathrm{km\,s^{-1}}$. The clustered accretion outbursts traced by the CO knots N1 to N4 exhibit time intervals on the order of tens of years. This timing of the four clustered accretion outbursts is consistent with the numerical prediction on the frequency of accretion outbursts triggered by disk gravitational instability and fragmentation \citep{2015ApJ...805..115V}. \citet{2019MNRAS.482.5459M} further report that minor to moderate accretion outbursts of tens of years triggered by disk gravitational instability and fragmentation are very frequent in the earliest stages of massive star formation.

\subsection{Episodic Mass Accretion and Ejection in the Formation Process of Massive Stars}
The SMA observations, conducted between 2008 and 2009 before the infrared outburst of S255IR~NIRS~3 in 2015, reveal collimated bipolar CO outflow from the same source \citep{2011A&A...527A..32W}. The recent high-resolution monitoring observations at multiple continuum bands by VLA and ALMA reveal the expansion of the radio jet triggered by the accretion outburst started in 2015 \citep{2018A&A...612A.103C,2023A&A...680A.110C}. Likewise, ALMA high-resolution observations reveal high-velocity north–south outflow traced by the CS $J=6-5,18-17$, and HDO $1_{1,1}-0_{0,0}$ transitions from NGC 6334I MM1 \citep{2018ApJ...866...87B,2018ApJ...863L..35M}. The red- and blueshifted gas from NGC 6334I MM1 is cospatial along the north–south direction. Large-scale northeast–southwest outflow detected by CO $J=4-3/9-8$ transition with APEX emerges at NGC 6334I MM1 \citep{2011ApJ...743L..25Q}. The north–south bipolar outflow from M17~MIR has well-separated red- and blueshifted emissions. The episodic ejecta of M17~MIR, revealed by the knotty structure of the blueshifted outflow, is the first clear case by which the accretion history of a forming massive protostar can be indirectly inferred.  

We estimate the masses of the CO knots N1 to N4 and S1, as listed in the third column of Table~\ref{tbl:2}. The details of this calculation are explained in Appendix~\ref{sec:appen_B}. The summed mass of knots N1 to N4 is $6.1\times10^{-2}\,M_\sun$, comparable to that ($5.1\times10^{-2}\,M_\sun$) of the redshifted knot S1. The comparable mass between the red- and blueshifted gas further suggests that the ejecta of M17~MIR might be symmetric on the red and blue sides.  

We assume that the mass ($M_\mathrm{out}$) of a CO knot accumulates through episodic mass ejection during contemporary clustered accretion outbursts and that the previously estimated $\Delta t$ well represents the duration of the clustered accretion outbursts. The mean outflow rate $\dot M_\mathrm{out}$ during each clustered accretion outburst is $M_\mathrm{out}/\Delta t$, as listed in the fifth column of Table~\ref{tbl:2}. Values of $\dot M_\mathrm{out}$ traced by knots N1 to N4 during the four clustered accretion outbursts range from $1.8\times10^{-4}\,M_\sun\,\mathrm{yr}^{-1}$ and $3.7\times10^{-4}\,M_\sun\,\mathrm{yr}^{-1}$. If considering the redshifted component ejected during the same period, values of $\dot M_\mathrm{out}$ are multiplied by a factor of 2, reaching between  $3.6\times10^{-4}\,M_\sun\,\mathrm{yr}^{-1}$ and $7.4\times10^{-4}\,M_\sun\,\mathrm{yr}^{-1}$.
The CO outflow traces the entrained gas rather than the ejecta directly from M17~MIR. The mass of the ejecta is roughly one-tenth that of the outflow \citep{2021MNRAS.500.3594R,2022MNRAS.510.2552R}. Applying the assumption of momentum conservation, a mass ratio of one-tenth can also be derived from the tangential mean velocity ($45\,\mathrm{km\,s^{-1}}$) of the CO knot N1/S1 and the tangential velocity of ejecta ($421\,\mathrm{km\,s^{-1}}$). With this mass ratio, the ejecta mass is one-tenth the mass of CO outflow. The mass ejection rate $\dot M_\mathrm{ejec}$ is reduced to the range from $3.6\times10^{-5}\,M_\sun\,\mathrm{yr}^{-1}$ to $7.4\times10^{-5}\,M_\sun\,\mathrm{yr}^{-1}$. The value of $\dot M_\mathrm{ejec}$ for the most recent clustered accretion outbursts, traced by the CO knot N1, is $\sim6.2\times10^{-5}\,M_\sun\,\mathrm{yr}^{-1}$.

The disk accretion rate $\dot M_\mathrm{acc}$ during the major accretion outburst around the 1990s is $\sim5\times10^{-3}\,M_\sun\,\mathrm{yr}^{-1}$, while that during the ongoing accretion outburst is $\sim1.7\times10^{-3}\,M_\sun\,\mathrm{yr}^{-1}$ \citep{2021ApJ...922...90C}. The minor and moderate accretion outbursts (e.g. the ongoing one of M17~MIR) might occur more frequently than major ones in the earliest stages of MYSOs \citep{2019MNRAS.482.5459M}. Therefore, taking the quiescent phase into account, we estimate a mean $\dot M_\mathrm{acc}\sim1\times10^{-3}\,M_\sun\,\mathrm{yr}^{-1}$ for the recent clustered accretion outbursts, which is lower than the value ($\sim1.7\times10^{-3}\,M_\sun\,\mathrm{yr}^{-1}$) presumed for the ongoing moderate accretion outburst.

The disk accretion and mass ejection processes in the earliest stage of massive protostars are crucial parts in understanding the formation mechanism of massive stars. Mass ejection, removing the excess angular moment of infalling material, is important for the growth of a massive protostar via disk accretion. We have independently estimated $\dot M_\mathrm{acc}$ and $\dot M_\mathrm{ejec}$ during the recent clustered accretion outbursts of M17~MIR. The value of $\dot M_\mathrm{ejec}/\dot M_\mathrm{acc}$ is $\sim6.2\%$, comparable to the ratio of the outbursting MYSO S255IR~NIRS~3 estimated from its radio jet \citep{2023A&A...680A.110C} and the ratios found for low- to intermediate-mass young stellar objects (YSOs) \citep{2013A&A...551A...5E,2016ARA&A..54..491B}. The tight connection between accretion and ejection found for M17~MIR in the earliest forming stage of an MYSO strongly supports the scenario that episodic mass accretion and ejection might be the potential way for the growth of a high-mass protostar from a lower-mass protostar.

The total mass of the CO knots is $\sim0.11\,M_\sun$. If the ratio $\dot M_\mathrm{ejec}/\dot M_\mathrm{acc}$ remains constant for the distant CO knots N2 to N4, the mass accumulated by frequent accretion outbursts over the past several hundred years amounts to  $\sim0.2\,M_\sun$. During the main accretion phase on the order of tens of thousands of years, the episodic accretion events would likely contribute more than about $10\,M_\sun$ to the final mass of M17~MIR, consistent with the prediction of the disk gravitational instability and fragmentation model \citep{2019MNRAS.482.5459M}.

\section{Summary and Conclusion} \label{sec:conclu}
The ALMA CO $J=2-1$ data reveal collimated and bipolar north–south outflow powered by M17~MIR. The blueshifted outflow shows four knotty structures N1 to N4 from near to far along the outflow axis, while the redshifted outflow appears as one knot S1. The EHV subknots of N1 and S1 are jetlike and contain subknot structures. The symmetrical morphology of the EHV subknots encourages us to assume that they trace the multiple ejecta triggered by the multiple accretion outbursts in the past decades reported in \citet{2021ApJ...922...90C}. This assumption aids in estimating a tangential velocity of $\sim421\,\mathrm{km\,s^{-1}}$ for the ejecta traced by the EHV subknots very close to the source. By assuming the same tangential velocity, we estimate the dynamical times of all the CO knots, which are in the range of $20-364$ yr. The four blueshifted CO knots imply four clustered accretion outbursts with timescales up to hundreds of years. The interval between the four clustered accretion outbursts and their duration is about tens of years, consistent with the prediction of the disk gravitational instability and fragmentation model. We also estimate the mass ejection rates, on the order of a few $10^{-5}\,M_\sun\,\mathrm{yr}^{-1}$, for the multiple ejecta triggered by the four clustered accretion outbursts. A ratio of ejection rate to accretion rate, that is, $\dot M_\mathrm{ejec}/\dot M_\mathrm{acc}$, is derived as $\sim6.2\%$ for the nearest CO knot N1. If this $\dot M_\mathrm{ejec}/\dot M_\mathrm{acc}$ value does not vary much during the main accretion phase of tens of thousands of years long, episodic accretion would contribute more than about $10\,M_\sun$ to the final mass of M17~MIR. The episodic accretion history of M17~MIR strongly supports the scenario that a massive star can be formed out from a lower-mass object by frequent episodic accretion events triggered by disk gravitational instability and fragmentation. The mass ejection phenomena accompanied with the accretion events may serve as effective diagnoses for the episodic accretion histories of forming MYSOs over a large time span. 

\begin{acknowledgments}

The authors appreciate the anonymous reviewer for reviewing this work. This work is supported by the National Natural Science Foundation of China (grant Nos. 12373030, U2031202). Z.C. acknowledges the Natural Science Foundation of Jiangsu Province (grants No. BK20231509). We would like to express our gratitude to Q. Zhang for the fruitful discussion that motivated our interest in this work. This paper makes use of the following ALMA data: ADS/JAO.ALMA\#2019.1.00994.S. ALMA is a partnership of ESO (representing its member states), NSF (USA) and NINS (Japan), together with NRC (Canada), MOST and ASIAA (Taiwan, China), and KASI (Republic of Korea), in cooperation with the Republic of Chile. The Joint ALMA Observatory is operated by ESO, AUI/NRAO and NAOJ.
\end{acknowledgments}

\vspace{5mm}
\software{CASA \citep{2007ASPC..376..127M}, Astropy \citep{2013A&A...558A..33A}, Matplotlib \citep{2007CSE.....9...90H}, Spectral-Cube \citep{2019zndo...3558614G}.}

\bibliography{ALMA_M17MIR}{}
\bibliographystyle{aasjournal}

\appendix

\section{Inclination angle of the outflow} \label{sec:appen_A}

The radiative transfer modeling for the spectral energy distribution (SED) of M17~MIR in two epochs, 2005 and 2017, returns some ``good-fit'' YSO models \citep[Table 9 in][]{2021ApJ...922...90C}. We adopt only the two ``good-fit'' YSO models with the smallest $\chi^2$ values for the 2017 SED, which is better constrained by the SOFIA/FORCAST observations at $19.7$ and $37.1\,\mu$m in 2017 \citep[see more details in][]{2021ApJ...922...90C}. The mean disk inclination angle for the two YSO models, weighted by $1/\chi^2$, is $70\degr$ with a standard deviation of $4\degr$. The nearly edge-on disk of M17~MIR could explain the large extinction ($A_V\sim 134$ mag) needed to fit the deep-absorption feature around $9.7\,\mu$m seen in the SEDs of M17~MIR. Assuming the outflow is perpendicular to the disk, which is inclined $70\degr$ to the plane of the sky, the inclination angle of the outflow is $20\pm4\degr$.

Redshifted CO emission is also observed at the position of the N2 knot in the channel maps and PV diagram of CO emission as mentioned before. The redshifted emission of N2 is much weaker than its blueshifted emission and is only visible at low velocity. We assume that the emission from N2 originates mostly from swept-up or vertical-splash low-velocity gas. The similar morphology of knot N2 in the blue- and redshifted channels suggests that the red- and blueshifted gases may originate from the same mass ejection events, which will facilitate the estimation of the outflow inclination. The projected offset along the outflow axis between the blue- and redshifted gas emission strongly depends on the inclination angle. We assume symmetrical blue- and redshifted components along the outflow axis. In Figure~\ref{fig:inclination}, we present a coarse geometric relation between the inclination angle \textit{i} relative to the plane of the sky, the axial offset \textit{d} between the blue and redshifted components and the spatial extension \textit{D} of N2 perpendicular to the outflow axis: $i=\arcsin{(d/D)}$. The blueshifted component of knot N2 emerges with knots N1 and N3, while the redshifted component is rather isolated in the $24-30\,\mathrm{km\,s^{-1}}$ channel in Figure~\ref{fig:ch_map}. From the redshifted component of knot N2 in the $24-30\,\mathrm{km\,s^{-1}}$ channel, we measure \textit{D} as $4\farcs1\pm0\farcs8$, where $4\farcs1$ is the average extension perpendicular to the outflow axis and $0\farcs8$ is the measurement uncertainty. The axial offset between the blue and redshifted components is denoted by \textit{d} in Figure~\ref{fig:pv_diagram}(a). We measure \textit{d} as $1\farcs1$ for knot N2. The inclination angle of the outflow is $\approx16\degr\pm3\degr$ by substituting the values of \textit{d} and \textit{D} into $i=\arcsin{(d/D)}$. This relatively small inclination angle ($\sim16\degr\pm3\degr$) agrees well with the value ($20\pm4\degr$) estimated from the disk inclination above.  The inclination angle of $16\degr$ estimated from the CO knot N2 is utilized in this work. The $3^\circ$ uncertainty may lead to approximately a 20\% underestimation or overestimation of the tangential velocities shown in Table~\ref{tbl:1}.

\setcounter{figure}{0}
\renewcommand{\figurename}{Figure}
\renewcommand{\thefigure}{A\arabic{figure}}

\begin{figure}
\centering
\includegraphics[width=0.5\textwidth]{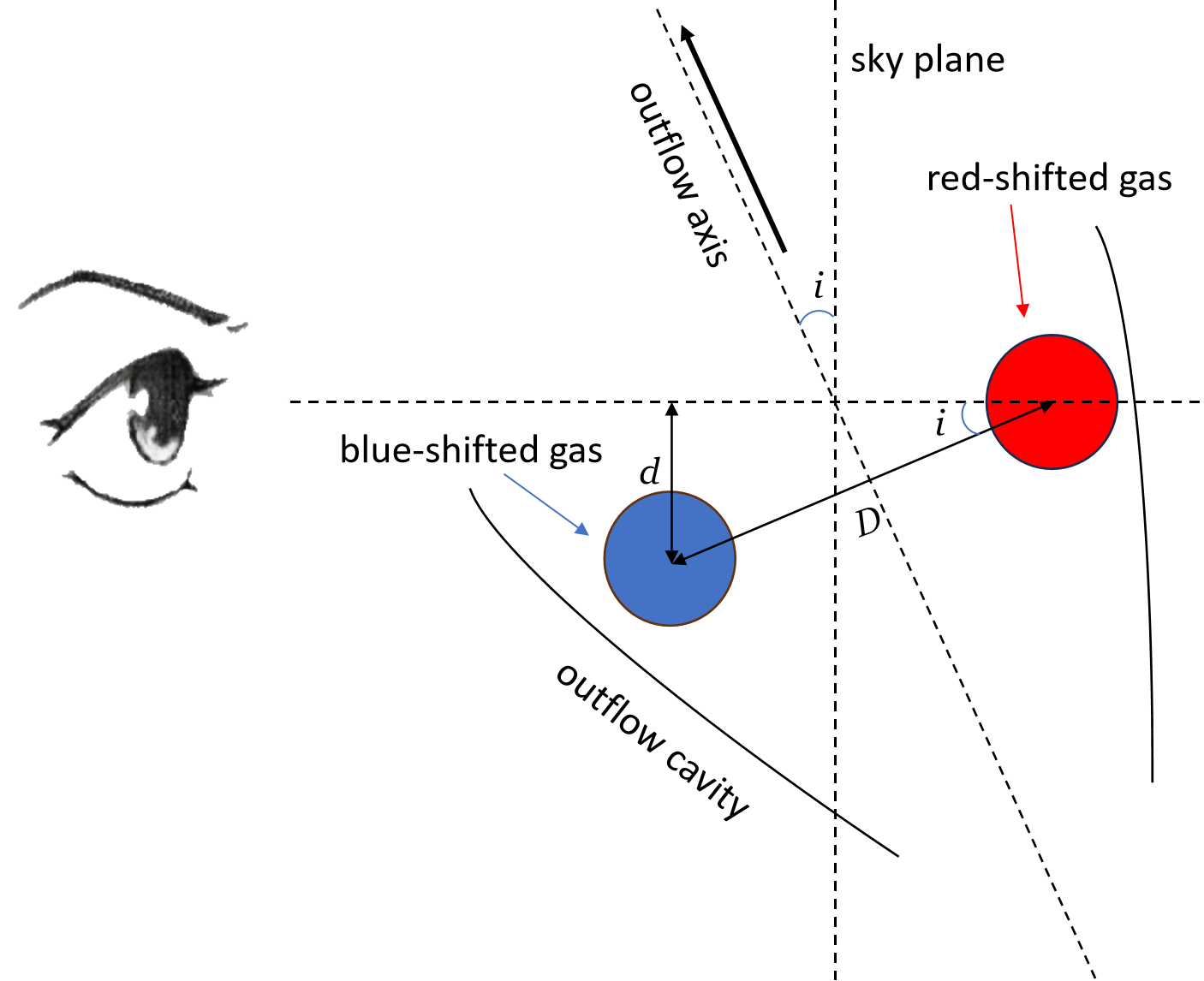}
\caption{The diagram illustrates the geometry of knot N2. The outflow orientation is bottom up, denoted by the black arrow. The blueshifted component is slightly closer to the central source than the redshifted component. \textit{d} represents this axial offset between them and is depicted in Figure~\ref{fig:pv_diagram}(a).
\textit{D} denotes the extension of knot N2 perpendicular to the outflow axis (the black horizontal line in the $24-30\,\mathrm{km\,s^{-1}}$ channel of Figure~\ref{fig:ch_map}).} 
\label{fig:inclination}
\end{figure}

\section{Outflow mass estimation} \label{sec:appen_B}

We utilize the methodology outlined in \citet{2015PASP..127..266M} and \citet{2019ApJ...886..130L} and assume a standard constant excitation temperature of 50\,K  to determine the CO column density of the outflow \citep{2010A&A...518L.121V,2014ApJ...783...29D,2015A&A...576A.109Y}. Following the optical-thin assumption, 
\[N_{\mathrm{CO}}(\mathrm{cm}^{-2})=1.086\times10^{13}(T_{\mathrm{ex}}+0.92)\mathrm{exp}\left(\frac{16.596}{T_{\mathrm{ex}}}\right)\int{T_{\mathrm{B}}dv}{,} \eqno(\mathrm{B.1})\]
where $N_\mathrm{CO}$ is the CO column density, $dv$ is the velocity interval in kilometers per second, $T_\mathrm{ex}$ is the line-excitation temperature, and $T_{\mathrm{B}}$ is brightness temperature in kelvin.
We can estimate the total gas mass employing
\[M_{\mathrm{outflow}}=\left[\frac{\mathrm{H_{2}}}{\mathrm{CO}}\right]d^2\overline{m}_{\mathrm{H_2}}\int{N_{\mathrm{CO}}(\Omega)d\Omega}{,} \eqno\mathrm{(B.2)}\] 
 where $\Omega$ is the total solid angle that the outflow subtends and $d\approx2.0$\,kpc is the distance to the source. We adopt the CO-to-H$_2$ abundance of $10^{-4}$ \citep{1987ApJ...315..621B} and the mean mass per hydrogen atom $\overline{m}_\mathrm{{H_2}}$ = 2.33. For each knot, we select the region within the lowest contour on both the blue- and redshifted sides in Figure~\ref{fig:pv_diagram}(b). We integrate emissions exceeding 7 $\mathrm{km\,s^{-1}}$ relative to the systemic velocity in the region to exclude the low-velocity emission from the cloud. We then recover the flux by assuming the knots extended as a 2D Gaussian distribution. Thirty percent of the mass for knots N1 and S1 were recovered, while 41\%, 57\%, and 49\% of the mass for N2, N3, and N4, respectively, were recovered. This ratio is determined by the ratio of each knot's peak integrated intensity to its cutoff threshold (value of the lowest contour in Figure~\ref{fig:pv_diagram}(b)).

\end{CJK*}
\end{document}